\documentclass[epjCONF]{svjour}
\usepackage{graphicx}
\usepackage[varg]{txfonts} 
\usepackage[latin1]{inputenc}
\session-title{The Space Photometry Revolution}
\begin{document}
\title{Connection between the period and
the amplitude of the Blazhko effect}
\author{Benk\H{o}, J\'ozsef M.\thanks{\email{benko@konkoly.hu}} 
\and Szab\'o, R\'obert}
\institute{Konkoly Observatory, MTA CSFK, Konkoly Thege u. 15-17.
H-1121 Budapest, Hungary}
\abstract{
We found a possible relationship between the
modulation period and the amplitude of the Blazhko
RR Lyrae stars: long modulation period
generally implies high modulation amplitude
while the short modulation period results in
small amplitude. Although this effect is much
more a tendency than a strict rule, it can be
detected easily in the space-born time series
data produced by Kepler and CoRoT. 
Good quality ground-based data show this relation, too.
This phenomenon could give us constraints for
the physics of the Blazhko effect.
} 
\maketitle
\section{Introduction}
\label{intro}\label{sec:1}
By investigating the Kepler Blazhko RR Lyrae light 
curves \cite{Ben2014} we found a possible
connection between the Blazhko period $P_{\mathrm B}$ and the
Fourier amplitude of the modulation frequency $A(f_{\mathrm B})$
($P_{\mathrm B}\propto A(f_{\mathrm B}$); see Fig.~9 in \cite{Ben2014}).
The literature was reviewed searching for this effect,
but nothing specific was found. Two papers were
published which implicitly suggest such a relation.
(i) Jurcsik, S\'odor \& V\'aradi \cite{Jur2005a} found a
correlation between the pulsation period $P_0$ and the
modulation amplitude of RR Lyrae stars ($P_0\propto A_2$).
Here the modulation amplitude $A_2$ was defined as
the sum of the Fourier amplitudes of the first four
modulation components:
$
A_2=A(f_0+f_{\mathrm B})+A(f_0-f_{\mathrm B})+
A(2f_0+f_{\mathrm B})+A(2f_0-f_{\mathrm B}).
$
(ii) Jurcsik et al. \cite{Jur2005b} found a correlation between
$P_0$ and the Blazhko period $P_{\mathrm B}$, as well ($P_0\propto P_{\mathrm B}$).

On the basis of these works we could deduce a
possible relation between the Blazhko period and
the modulation strength: viz.
if $P_0\propto A_2$ and $P_0\propto P_{\mathrm B}$ then 
$P_{\mathrm B}\propto A_2$.
What is the connection between the parameter $A_2$
and $A(f_{\mathrm B})$? As we showed in a simplified
mathematical framework \cite{Ben2011}, 
the modulation component amplitudes in $A_2$
depend on the strength of the frequency modulation
(FM), while $A(f_{\mathrm B})$ depends on the amplitude
modulation (AM) only.
Therefore, using $A(f_{\mathrm B})$ is clearly superior in
characterizing AM. There is, however, a drawback.
The number of stars of known $A(f_{\mathrm B})$ is
small, because $f_{\mathrm B}$ can be detected only in space-born
and the best quality ground-based data.

\section{Sample and Method}\label{sec:2}
We collected stars where $A(f_{\mathrm B})$ are
known. These stars were observed from
space by Kepler \cite{Ben2014}, by
CoRoT \cite{Cha2009}, \cite{Por2010}, \cite{Gug2011}, and
from the ground by the Konkoly Blazhko
Survey (KBS, \cite{Jur2009}).
These different observations were
obtained in different color bands (e.g. {\it V,
R, I, K$_{\mathrm p}$}), hampering their uniform
handling. The spectral response
function of the CoRoT and Kepler
detectors are similar, but the CoRoT
band is a bit wider than Kepler's one
\cite{Auv2009}, \cite{Van2009}, therefore, we have to scale the
CoRoT amplitudes. Nemec at al. \cite{Nem2011} found
empirical transformations between
amplitudes (e.g. $A_{\mathrm{tot}}$, $A_1$) in bands {\it K$_{\mathrm p}$}\/ and
{\it V}\/. The transformed KBS data, however,
do not show a clear correlation.
There is an alternate parameter for
characterizing the strength of AM: the
amplitude of the envelope curve $A_{\mathrm{max}}$.
This parameter can be determined
easily, but it depends also on the
pulsation amplitude. We checked the
ratio on KBS data using this $A_{\mathrm{max}}$ values
and found a tight correlation between $P_{\mathrm B}$
and $A(V)_{\mathrm{max}}$ (see blue symbols in Fig.~\ref{figBen:1}).
We applied the empirical transformation
formula of \cite{Nem2011} between
the total amplitude in bands {\it K$_{\mathrm p}$} and {\it V}\/. 
If we plot the result in Fig.~\ref{figBen:1}. (red circles), 
Kepler stars show similar
correlation than KBS stars. This 
$P_{\mathrm B}\propto A_{\mathrm{max}}$ ratio could be tested in the future
by using massive photometric data
basis (e.g. MACHO, OGLE).

\section{Interpretation}
\begin{figure}
  \includegraphics[width=13.5cm]{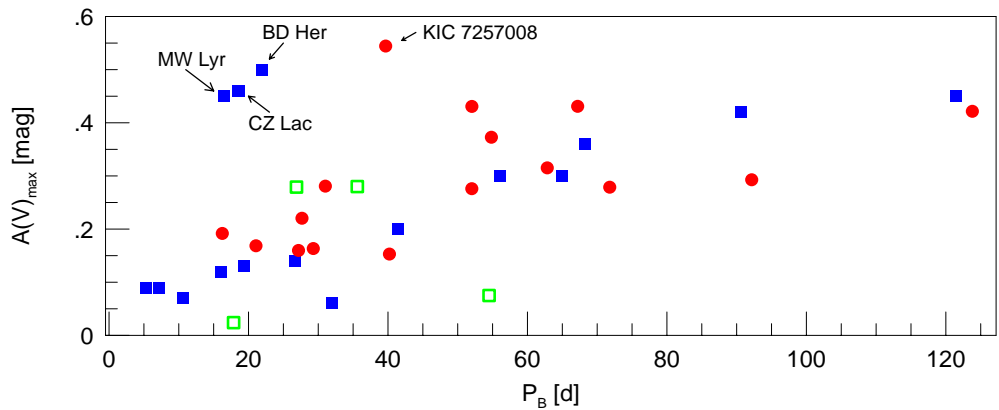}
\caption{Blazhko period $P_{\mathrm{B}}$ vs. total {\it V}\/ amplitude 
of the maxima (envelope) curve. Red and blue symbols denote
Kepler and Galactic field (Konkoly Blazhko Survey) RR Lyrae 
stars, respectively. Open (green) squares indicate transformed values of 
the CoRoT stars. The typical errors are smaller than the symbols.
}
\label{figBen:1}       
\end{figure}

Similar effect is common in hydrodynamical systems: e.g. weakly
dissipating systems forced to show high amplitude by long time-scale
perturbing forces only \cite{Mol2012}.
We call the reader's attention to the
deviating stars (CZ\,Lac, MW\,Lyr, BD\,Her and KIC\,7257008 
in Fig.~\ref{figBen:1} and V355\,Lyr in Fig.~9 of \cite{Ben2014}). 
These may represent a separate group
within Blazhko RR Lyrae stars. The
existence of this group can also be tested
by using large data bases.
\begin{acknowledgement}
The research leading to these results has received funding from the
European Community's Seventh Framework Programme (FP7/2007-2013)
under grant agreements no. 269194 (IRSES/ASK) and no. 312844
(SPACEINN). This work was also supported by the following grants: ESA
PECS No 4000103541/11/NL/KML, No 4000110889/\-14/\-NL/\-NDe and the
Hungarian OTKA Grant K-83790.
\end{acknowledgement}

\end{document}